\documentclass[a4paper,11pt]{article}
\pdfoutput=1 
\usepackage[utf8]{inputenc}

\usepackage{jinstpub}

\usepackage{graphicx}
\usepackage[T1]{fontenc} 
\usepackage{graphbox}
\usepackage{lineno}
\usepackage{placeins}
\usepackage{feynmf}
\usepackage{xspace}

\newcommand{\ie}{{\sl i.e.}~}
\newcommand{\eg}{{\sl e.g.}~}

\newcommand{\pt}{\ensuremath{p_{\rm T}}}
\newcommand{\px}{\ensuremath{p_{\rm x}}}
\newcommand{\py}{\ensuremath{p_{\rm y}}}
\newcommand{\pz}{\ensuremath{p_{\rm z}}}
\newcommand{\etmiss}{\ensuremath{E_{T}^{\rm miss}}}
\newcommand{\ephimiss}{\ensuremath{E_{\phi}^{\rm miss}}}

\newcommand*{\Wboson}{\ensuremath{W}\xspace}

\newcommand{\ttbar}{\ensuremath{t\bar{t}}}
\newcommand{\MadGraph}{\textsc{MadGraph5}}
\newcommand{\Pythia}{\textsc{Pythia8}}

\newcommand{\Delphes}{\textsc{Delphes3}}

\newcommand{\FastJet}{\textsc{FastJet}}
\newcommand{\Keras}{\textsc{Keras}}
\newcommand{\Tensorflow}{\textsc{Tensorflow}}
\newcommand{\Hyperopt}{\textsc{Hyperopt}}
\newcommand{\Tune}{\textsc{Tune}}
\newcommand{\AngryTops}{\textsc{AngryTops}}

\newcommand{\chisqfit}{\ensuremath{\chi^2}-\textsc{fit}}
\newcommand{\KLFitter}{\textsc{KLFitter}}
\newcommand{\PseudoTop}{\textsc{PseudoTop}}

\newcommand{\GeV}{\ensuremath{\text{Ge\kern -0.1em V}}}

\title{\boldmath
Bidirectional Long Short-Term Memory (BLSTM) neural networks for reconstruction of top-quark pair decay kinematics 
}

\begin{document}

\author{Fardin Syed},
\author{and Riccardo Di Sipio},
\author{and Pekka K. Sinervo, C.M.}
\affiliation{Department of Physics, University of Toronto, 60 St George St, Toronto, ON M5S 1A7, Canada}

\emailAdd{riccardo.disipio@utoronto.ca}


\abstract{
A probabilistic reconstruction using machine-learning of the decay kinematics of top-quark pairs produced in high-energy proton-proton collisions is presented. A deep neural network whose core consists of a Bidirectional Long Short-Term Memory (BLSTM) is trained to infer the four-momenta of the two top quarks produced in the hard scattering process. The \MadGraph{}+\Pythia{} Monte Carlo event generator is used to create a sample of top-quark pairs decaying in the $\mu$+jets channel, whose final-state objects are used to create the input to the deep neural network. Distortions due to limited resolution of the experimental apparatus are simulated with the \Delphes{} fast detector simulator. The level of agreement between the Monte Carlo predictions and the BLSTM for kinematic distributions at parton level is comparable to that obtained using a benchmark method that finds the jet permutation that minimizes an objective function. The code is publicly available on the repository \url{https://github.com/IMFardz/AngryTops} .
}


\maketitle
\flushbottom

\section{Introduction}
Studies of energetic top quarks produced in hadron collisions provide a unique window into our theoretical framework for particle physics, known as the Standard Model \cite{Quadt:2007jk}.   They are also important for probing physics beyond the Standard Model. As the most massive fundamental particle with a rest mass of $\approx$ 173~GeV, the top quark can only be observed by its decay products and their corresponding signatures in detectors operating at particle colliders like the Large Hadron Collider (LHC). 

The production and decay of a top-quark pair results in six partons in the final state, including charged and neutral leptons and jets of particles arising from the daughter quarks and gluons.  The complexity of the resulting events and the finite resolution of the detectors makes the challenge of reconstructing the top-quark momenta exceptionally difficult. Improvements in top-quark reconstruction are essential for understanding rare processes and making precision measurements of top-quark cross sections. Current reconstruction routines such as \KLFitter{} \cite{KLFitter} and \PseudoTop{} \cite{PseudoTops} attempt to solve this problem algorithmically or by fitting a likelihood function, respectively. 
Both can be considered as improvements on a more basic algorithm, known as \chisqfit~\cite{Abulencia:2005aj}, which attempts to reconstruct the decay chains by assigning one jet uniquely to each outgoing parton. The permutation that minimizes an objective function is used to define the assignment. However, if the jet produced by any of these partons is lost because of limited acceptance and resolution, the reconstruction of the event kinematics is compromised. Also, the jets arising from the top-quark decay daughters, which are defined as clusters of the observed constituents (\eg calorimeter cells or tracks in the inner detector), are selected by applying a cut on their transverse momentum (\pt).\footnote{
We use a right-handed coordinate system with its origin at the nominal $pp$\ interaction point in the center of the detector and the $z$-axis along the beam pipe. 
The $x$-axis points from the centre of the detector to the center of the LHC ring and the $y$-axis points upward. 
Cylindrical coordinates $(r,\phi)$ are used in the transverse plane, $\phi$ being the azimuthal angle around the $z$-axis. 
The pseudorapidity is defined in terms of the polar angle $\theta$ as $\eta=-\ln\tan(\theta/2)$.
}\ 
Because of this reason, it is fundamentally not possible to identify quark jets with a \pt~smaller than the threshold, unless some extrapolation is applied. 
In principle, a probabilistic approach as the one presented in this work does not suffer from the limitations of \chisqfit, and is able to perform such extrapolation without any need to rely explicitly on parametric transfer functions as in the case of \KLFitter.

The aim of this paper is to introduce a machine-learning  approach to top quark reconstruction. In our analysis, we train \AngryTops{}, our machine learning software package, to reconstruct top-quark, bottom-quark and \Wboson-boson four-momenta in the lepton+jets \ttbar~decay channel in $pp$\ collisions at 13~TeV. 
In this topology, one top quark has decayed fully hadronically while
the other top quark has decayed semileptonically.  
We then evaluate \AngryTops{} by comparing its performance to \chisqfit.

In section \ref{Monte Carlo Sample}, we outline our procedure for generating Monte Carlo (MC) events for the training and testing of our networks. Then, in Section \ref{Network Architecture}, we describe the network architecture that lead to the best performance during our study. Section \ref{Training} describes in further detail the training procedure for our network architectures. In Section \ref{Results} we evaluate our model by comparing it against \chisqfit. Finally, in Sections \ref{Observations} and \ref{Conclusions} we summarize our findings and discuss the future of machine learning in the kinematic reconstruction of top-quark momenta.

\section{Monte Carlo Sample} \label{Monte Carlo Sample}
\begin{figure}[htbp]
\centering
\includegraphics[scale=1.0]{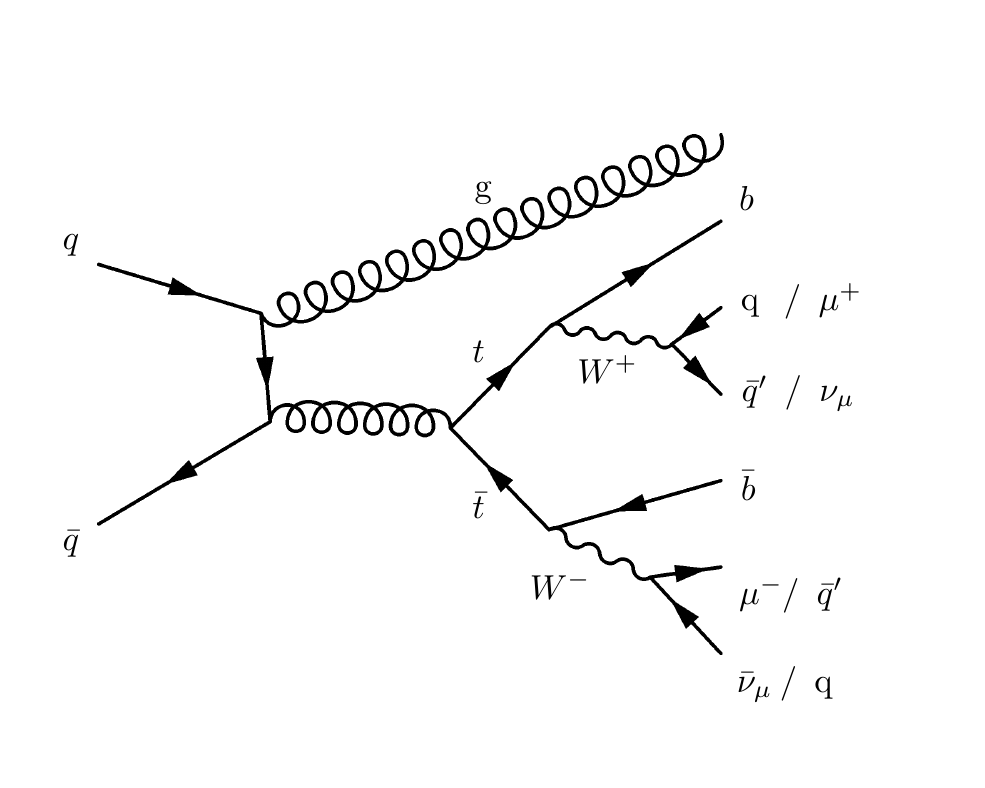}
\caption{\small{Example of a leading-order \ttbar{} production Feynman diagram used to train the BLSTM network. The final state shown consists of a muon, a neutrino and at least four jets. 
Up to one additional outgoing quark or gluon is considered in the matrix element calculation. Matching partons arising in the matrix-element calculation with those produced in the subsequent parton showering model is performed by the MLM matching algorithm \cite{mlm_matching}.}}
\label{fig:ttbar_diagram}
\end{figure}

To train \AngryTops{}, a sample of 200 million \ttbar{} events has been created. The \MadGraph{} Monte Carlo (MC) event generator \cite{MadGraph} has been used to calculate the amplitudes of the leading-order process $pp\rightarrow \ttbar$ with up to one additional outgoing quark or gluon, as shown in Fig.\ref{fig:ttbar_diagram}. The \Pythia{} generator \cite{Pythia8} was used to carry out the parton-showering of quarks and gluons and MLM matching \cite{mlm_matching}\ was employed to model how the parton showers were matched to the \MadGraph{}\ matrix-element calculcation. Finally, the detector simulation \Delphes{} has been used to simulate the effect of detector response. An average of 25 additional soft-QCD $pp$ collisions (pile-up) were overlaid to reproduce realistic data-taking conditions at the LHC.

In what follows, we will refer to the hadronically and semileptonically 
decaying top quarks as the ``hadronic top quark'' and the ``semileptonic top quark,'' respectively.

Electrons, muons, jets and missing transverse energy are reconstructed by \Delphes{} algorithms. Jets were reconstructed using the anti-$k_T$ algorithm \cite{antikt} as implemented in \FastJet{} \cite{fastjet}, with a distance parameter $R$ = 0.4.
We only consider the semileptonic top-quark decay modes that result in an energetic electron or muon and its associated neutrino.
In addition, we require that the transverse momenta and pseudorapidity of the muon, \Wboson~bosons and $b$ quarks are greater than $20$ GeV and less than 2.5, respectively. After these additional cuts, we are left with roughly 5 million events for training and testing our BLSTMs. 

\section{Network Architecture} \label{Network Architecture}
The input for \AngryTops{} is a 36 element array, which is reshaped into a (6 x 6) matrix in the first network layer. The first six elements of the input correspond to the following: (muon $\px$, $\py$, $\pz$, muon arrival time of flight $T_0$, missing transverse energy \etmiss, missing energy azimuthal angle \ephimiss). The subsequent five columns in the input matrix each correspond to an input jet and are defined as follows (jet \px, \py, \pz, energy $E$, mass $M$, $b$-tagging state $B$), where the $b$-tagging state is either 0 (not-tagged) or 1 (tagged). In the case when there are only 4 jets present in the event, the last jet column is set to all zeros. Our matrix of inputs is written as
\begin{eqnarray}
    \begin{pmatrix}
    \px^{\mu} & \px^{\rm{j,1}} & \px^{\rm{j,2}} & \px^{\rm{j,3}} & \px^{\rm{j,4}} & \px^{\rm{j,5}} \\
    \py^{\mu} & \py^{\rm{j,1}} & \py^{\rm{j,2}} & \py^{\rm{j,3}} & \py^{\rm{j,4}} & \py^{\rm{j,5}} \\
    \pz^{\mu} & \pz^{\rm{j,1}} & \pz^{\rm{j,2}} & \pz^{\rm{j,3}} & \pz^{\rm{j,4}} & \pz^{\rm{j,5}} \\
    T_0^{\mu} & E^{\rm{j,1}} & E^{\rm{j,2}} & E^{\rm{j,3}} & E^{\rm{j,4}} & E^{\rm{j,5}} \\
    \etmiss   & M^{\rm{j,1}} & M^{\rm{j,2}} & M^{\rm{j,3}} & M^{\rm{j,4}} & M^{\rm{j,5}} \\
    \ephimiss & B^{\rm{j,1}} & B^{\rm{j,2}} & B^{\rm{j,3}} & B^{\rm{j,4}} & B^{\rm{j,5}}
    \end{pmatrix}.
\end{eqnarray}

The output for our model is a (6 x 3) matrix, where each row corresponds to $\px$,  $\py$ and $\pz$ for the  bottom quark from the hadronic top-quark decay, the bottom quark from the semileptonic top-quark decay, hadronic \Wboson~boson, leptonic \Wboson~boson, hadronically decaying and semileptonically decaying top quark, respectively. For the purpose of this analysis, we fix the top-quark mass to $172.5$~\GeV and that of the \Wboson~boson to $80.4$~\GeV. The output matrix is
\begin{eqnarray}
    \begin{pmatrix}
    \px^{b,\rm{had}} & \py^{b,\rm{had}}   & \pz^{b,\rm{had}} \\
    \px^{b,\rm{lep}} & \py^{b,\rm{lep}}   & \pz^{b,\rm{lep}} \\
    \px^{W,\rm{had}} & \py^{W,\rm{had}}   & \pz^{W,\rm{had}} \\
    \px^{W,\rm{lep}} & \py^{W,\rm{lep}}   & \pz^{W,\rm{lep}} \\
    \px^{t,\rm{had}} & \py^{t,\rm{had}}   & \pz^{t,\rm{had}} \\
    \px^{t,\rm{lep}} & \py^{t,\rm{lep}}   & \pz^{t,\rm{lep}} \\
    \end{pmatrix}.
\end{eqnarray}

\begin{figure}[htbp]
\centering
\includegraphics[scale=0.35]{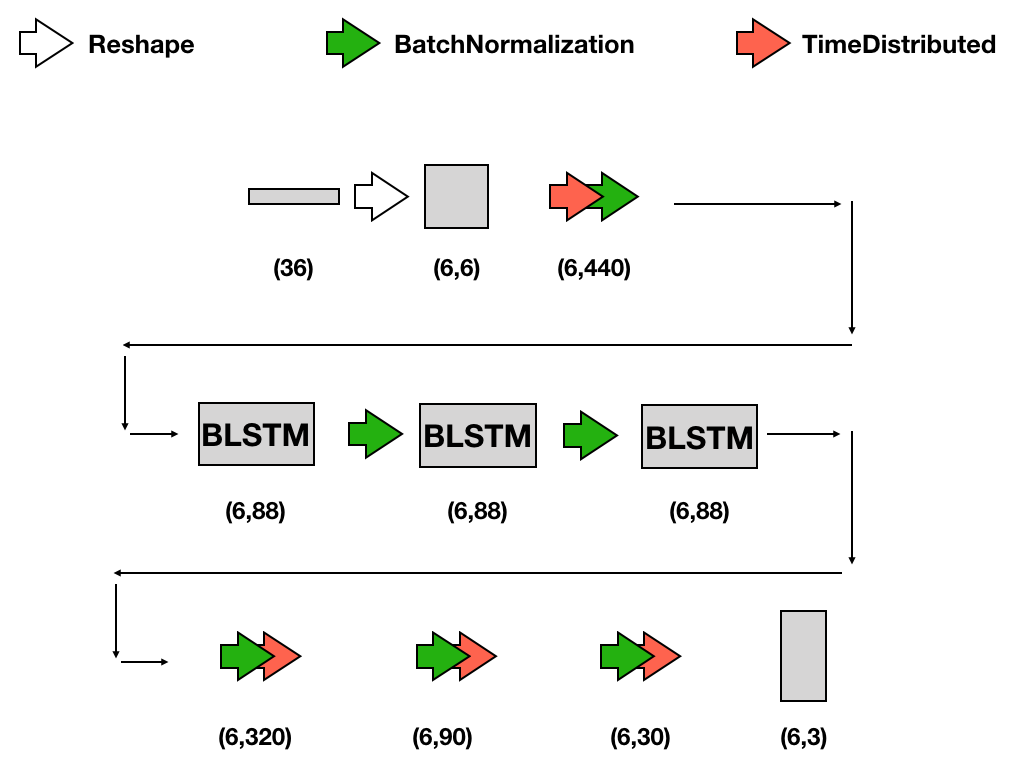}
\caption{\small{A diagram of the BLSTM network architecture highlights employed in this study.}}
\label{fig:model_diagram}
\end{figure}

Our network consists of 329,913 trainable parameters and 17 different layers. We experimented with Convolutional Neural Networks (CNNs), Long Short Term Memory (LSTMs), Feed Forward Neural Networks (FFNNs) and Bidirectional Long Short Term Memory (BLSTMs).  As described below, we found that BLSTMs performed the best. A diagram of our network architecture can be seen in Figure \ref{fig:model_diagram}. Our code is written with \Keras \cite{Keras} with a \Tensorflow{} 2.0 RC \cite{tensorflow} back-end. 

\section{Training} \label{Training}
For network training, we select an Adam Optimizer~\cite{kingma2014adam}\ with a learning rate of $10^{-4}$ and a Mean Squared Error (MSE) loss function. 

For each choice of network architecture, we scan through the hyper-parameter space and select the draw which leads to the lowest loss function. With \Hyperopt \cite{Hyperopt}, we used a uniform distribution to select the size of the network layers and network activation functions. With \Tune \cite{Tune}, we use an Asynchronous HyperBand Scheduler to train 1000 different draws from the hyper-parameter space in simultaneous batches of 8.

Of the five million selected events from the Monte-Carlo simulation, we use 90 \% for training and the other 10 \% for testing. For each training epoch, we set asid an additional 10 \% of the training set  for validation. A model's training is stopped when its validation loss function begins to increase. 

Before the start of training, we scale all the network input and outputs with a MinMax Scaling, which sets the minimum and maximum value for each kinematic variable to -1 and 1. We have also experimented with shuffling the training set, ordering the inputs by different variables, and scaling to Gaussian parameter distributions with mean 0 and a variance of 1. We did not find any significant improvements when using these alternative scaling techniques. 

\section{Results} \label{Results}
The output of the deep neural network is compared to a benchmark reconstruction method inspired by \chisqfit, \ie based on finding the jet permutation that minimizes the objective function
\begin{eqnarray}
\chi^2 =
    \frac{(m_{jjb}-m_t^{MC})^2}{\sigma_t^2} +
    \frac{(m_{jj}-m_W^{MC})^2}{\sigma_W^2} +
    \frac{(m_{l\nu b}-m_t^{MC})^2}{\sigma_t^2} +
    \frac{(m_{l\nu}-m_W^{MC})^2}{\sigma_W^2},
\end{eqnarray}
where $m_t^{MC} = 172.5$~\GeV, $\sigma_t = 30$~\GeV, 
$m_W^{MC} = 80.4$~\GeV and $\sigma_W = 20$~\GeV. 
In order to perform the calculation, up to the first five jets (ordered 
by decreasing transverse momentum) are considered. 
Then, each permutation consists of four jets that are uniquely assigned to the hadronically decaying top quark, hadronically decaying \Wboson~boson and semi-leptonically decaying top quark. 
Information about $b$-tagging and lepton arrival time of flight are 
not considered. 
The masses of the top quark, \Wboson~boson and bottom quark are used to calculate the energy component of the associated four-momenta. 
The neutrino four-momentum component along the $z$ axis (\pz) is estimated from the missing transverse energy and the quadratic \Wboson-boson mass constraint as in \cite{PseudoTops}. In the case of degenerate solutions, the smallest \pz{} value is selected.

\begin{figure}[htbp]
\centering
\includegraphics[scale=0.25]{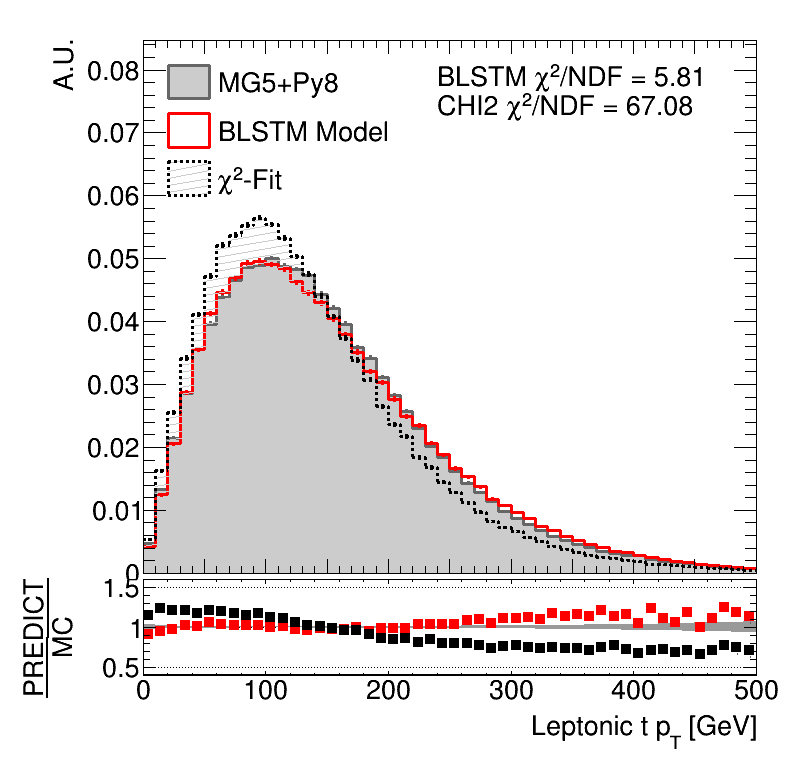}
\includegraphics[scale=0.25]{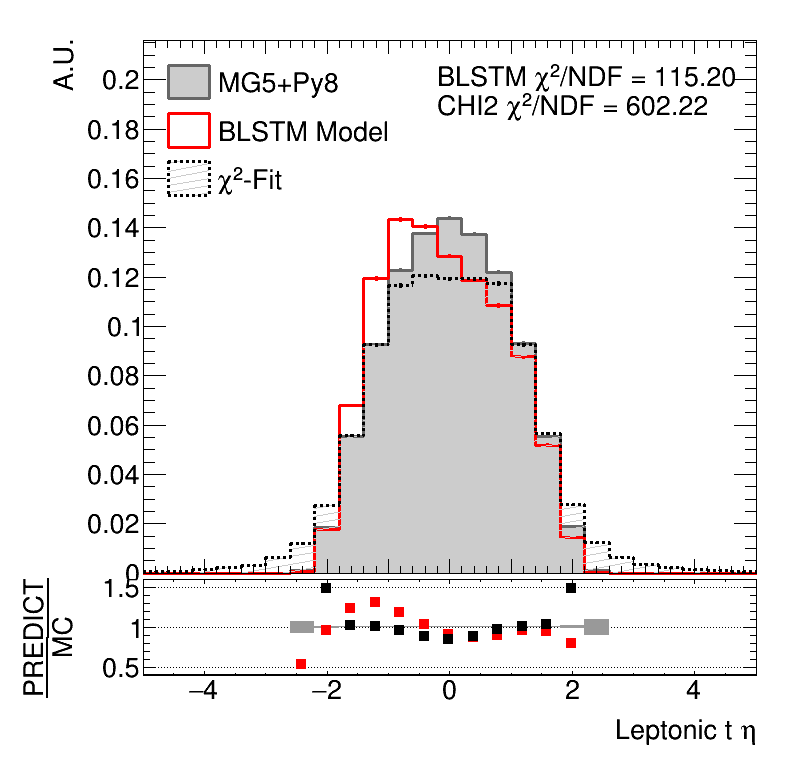}
\includegraphics[scale=0.25]{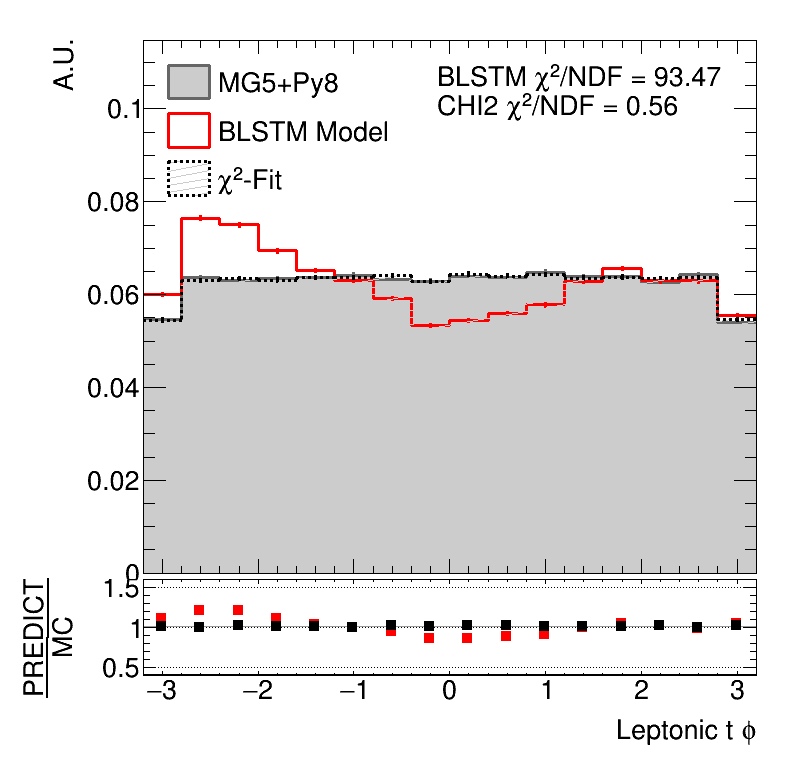}
\includegraphics[scale=0.25]{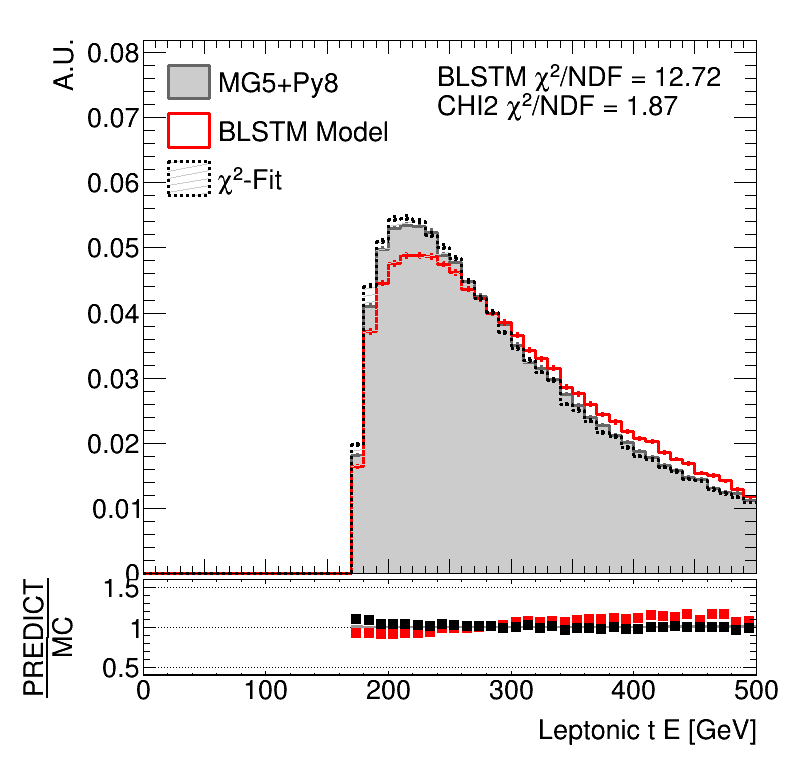}
\caption{\small{Reconstructed semileptonic top quark observables. The gray filled area represents the  prediction obtained using the \MadGraph{}+\Pythia{}~Monte Carlo event generator. 
The black dashed line is obtained from the permutation of jets which minimizes the $\chi^2$. The red solid line is the output of the BLSTM.}}
\label{fig:results:t_lep}
\end{figure}

\begin{figure}[htbp]
\centering
\includegraphics[scale=0.25]{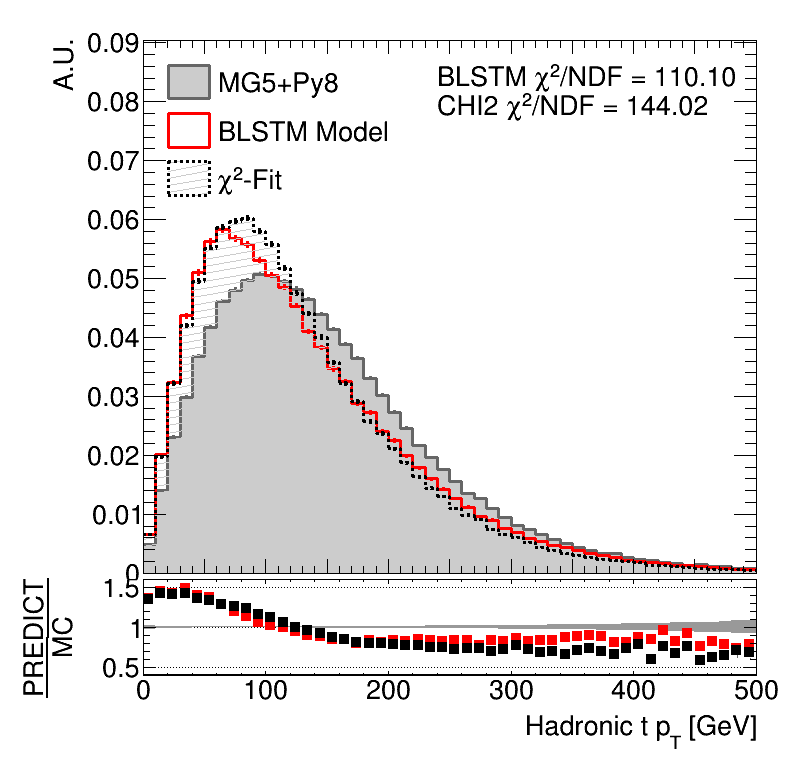}
\includegraphics[scale=0.25]{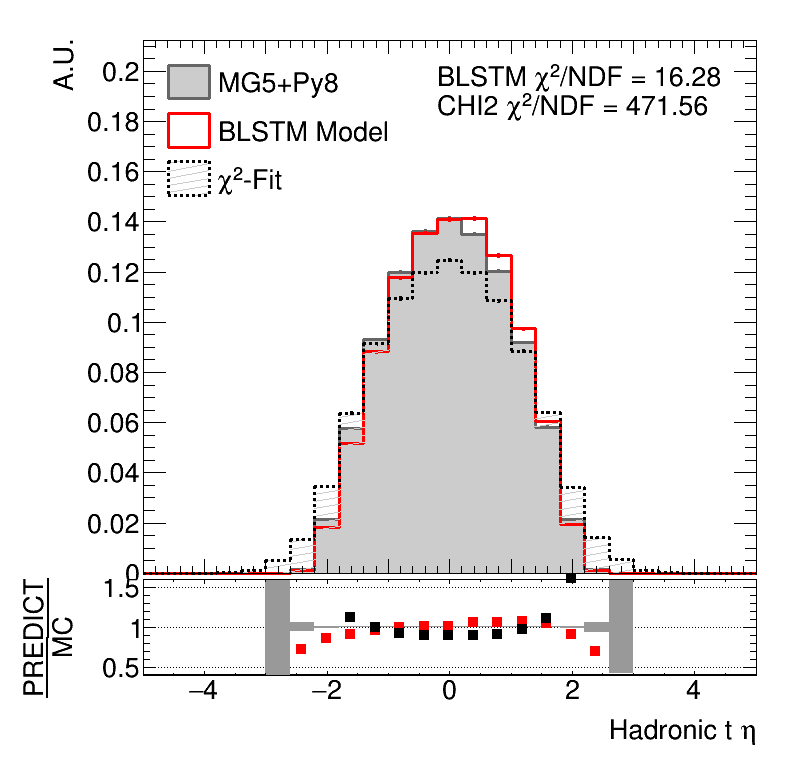}
\includegraphics[scale=0.25]{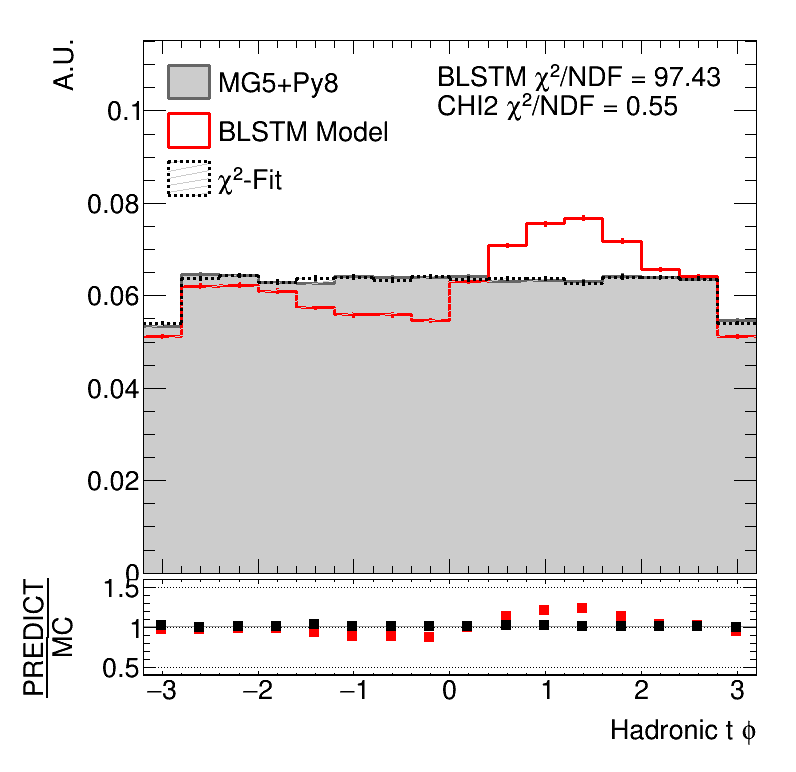}
\includegraphics[scale=0.25]{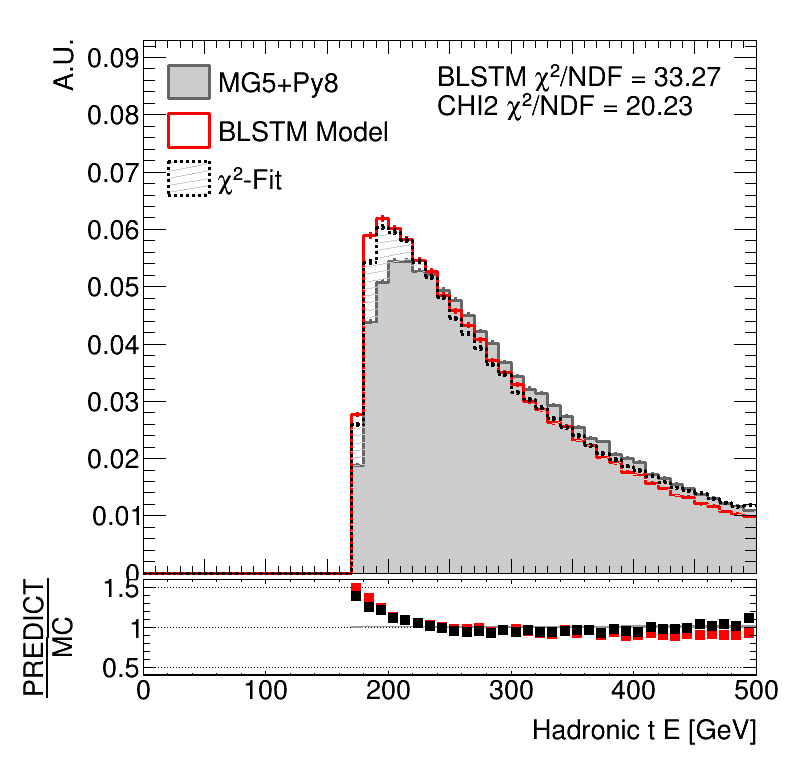}
\caption{\small{Reconstructed hadronic top quark observables. The gray filled area represents the  prediction obtained using the \MadGraph{}+\Pythia{}~Monte Carlo event generator. The black dashed line is obtained from the permutation of jets which minimizes the $\chi^2$. The red solid line is the output of the BLSTM.}}
\label{fig:results:t_had}
\end{figure}

\begin{figure}[htbp]
\centering
\includegraphics[scale=0.25]{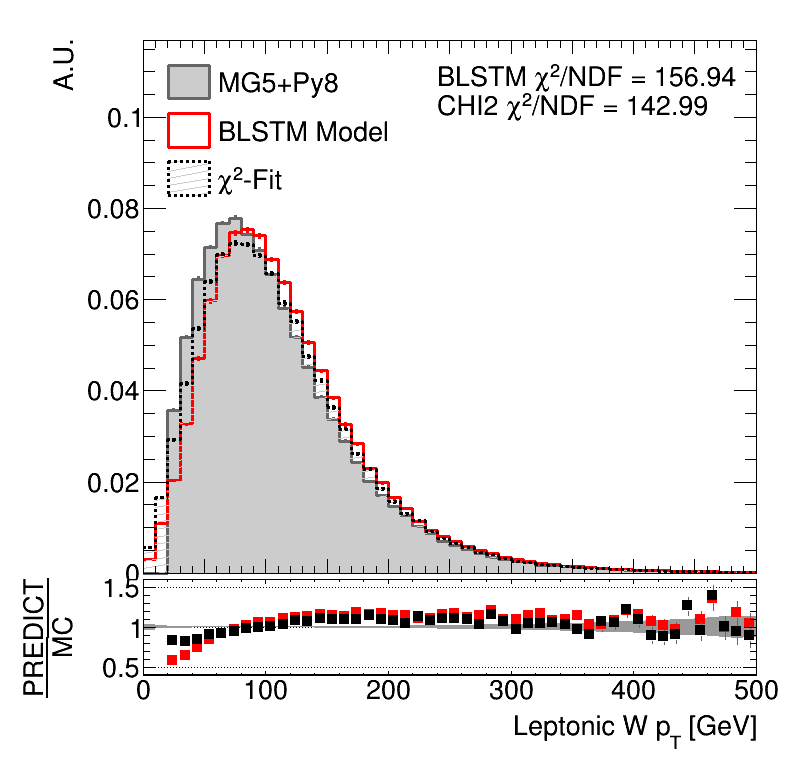}
\includegraphics[scale=0.25]{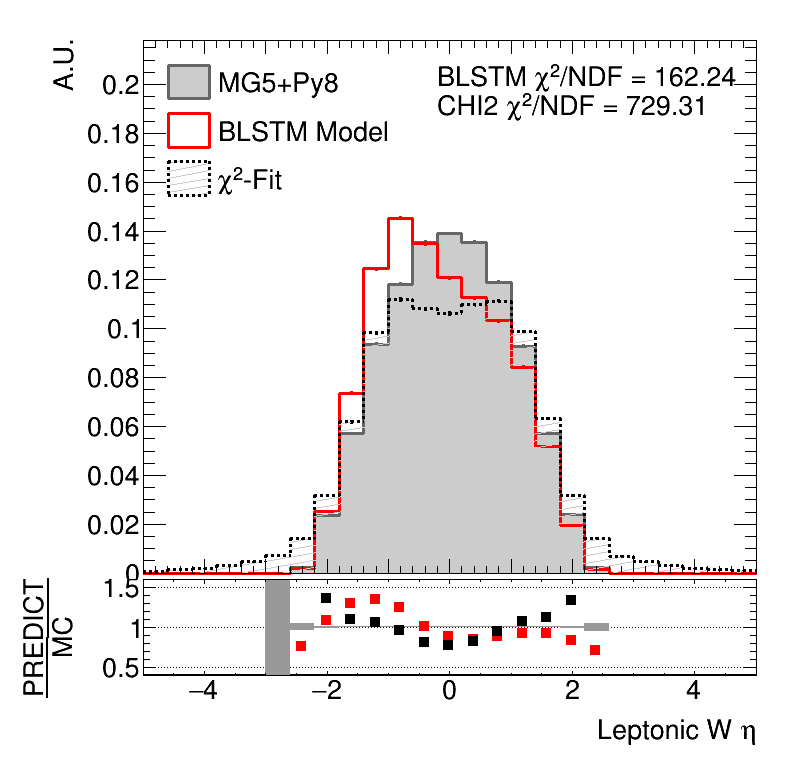}
\includegraphics[scale=0.25]{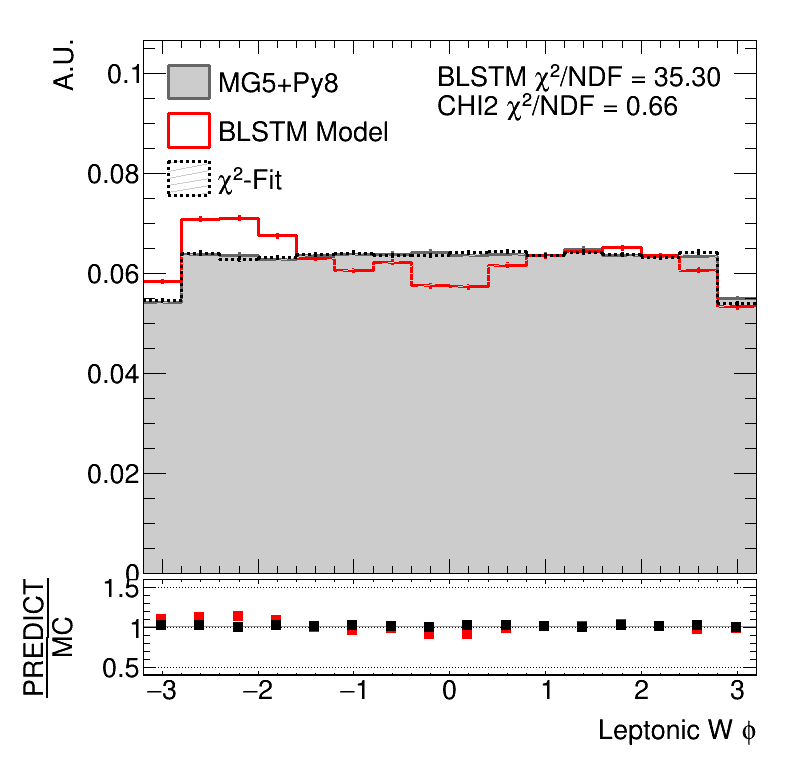}
\includegraphics[scale=0.25]{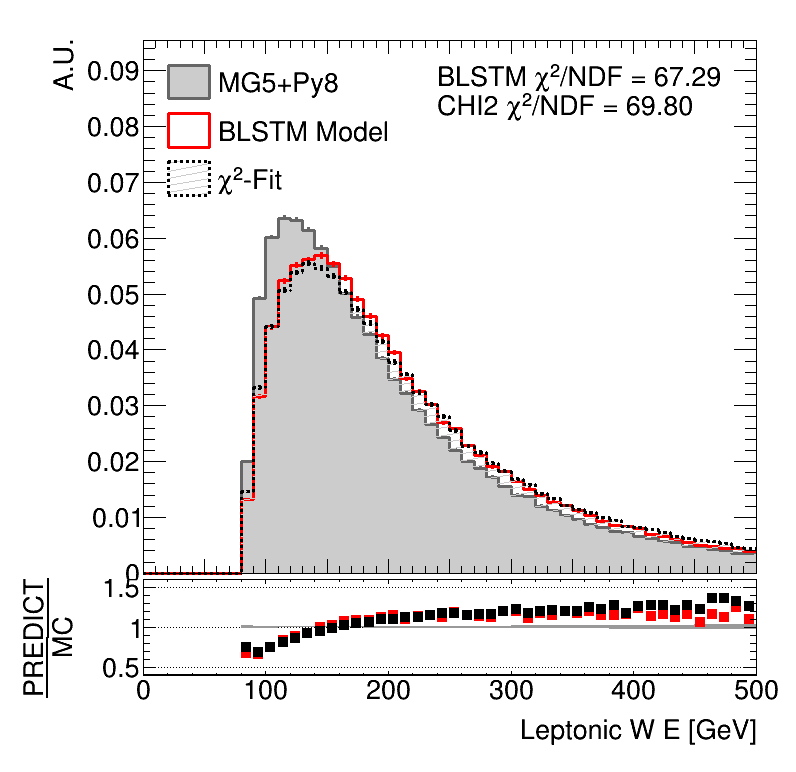}
\caption{\small{Reconstructed \Wboson~boson observables for the semileptonic top quark. The gray filled area represents the  prediction obtained using the \MadGraph{}+\Pythia{}~Monte Carlo event generator. The black dashed line is obtained from the permutation of jets which minimizes the $\chi^2$. The red solid line is the output of the BLSTM.}}
\label{fig:results:W_lep}
\end{figure}

\begin{figure}[htbp]
\centering
\includegraphics[scale=0.25]{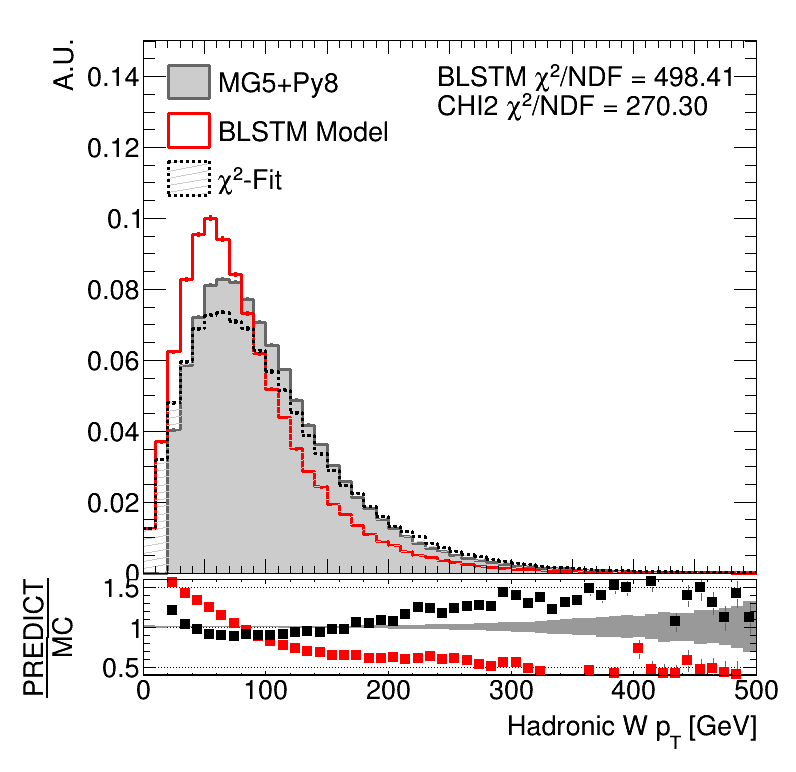}
\includegraphics[scale=0.25]{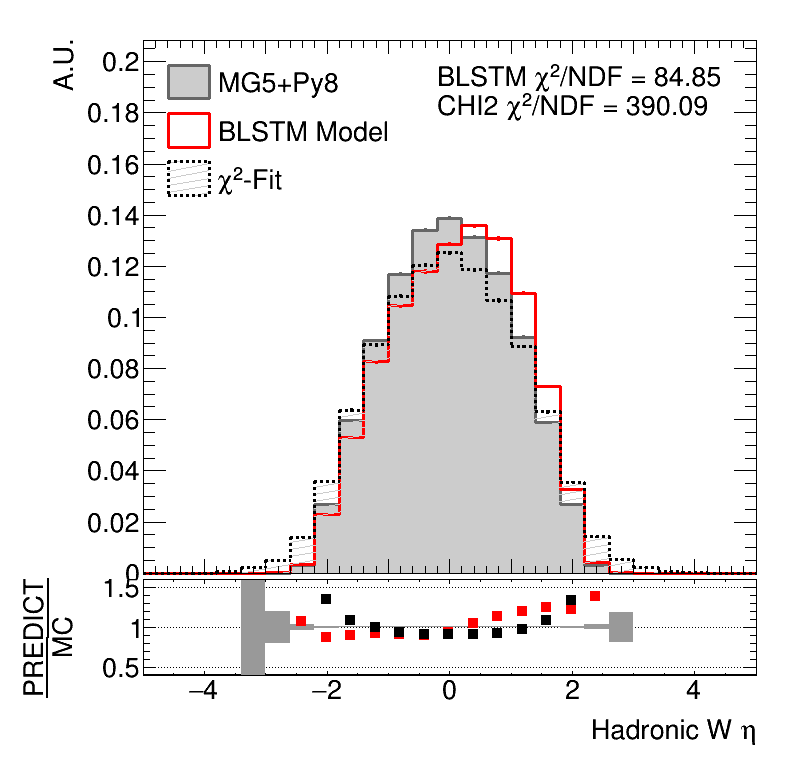}
\includegraphics[scale=0.25]{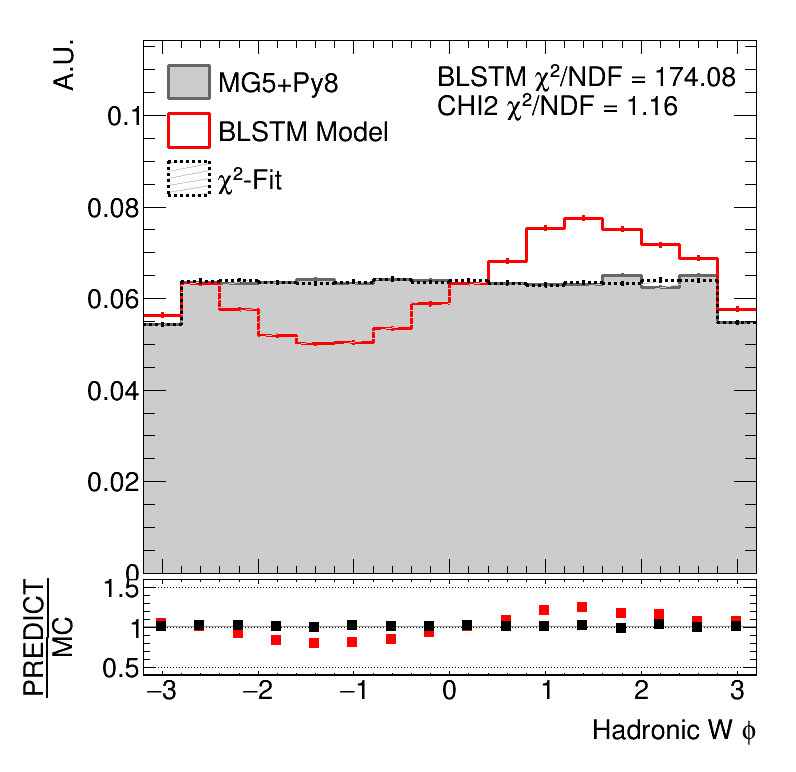}
\includegraphics[scale=0.25]{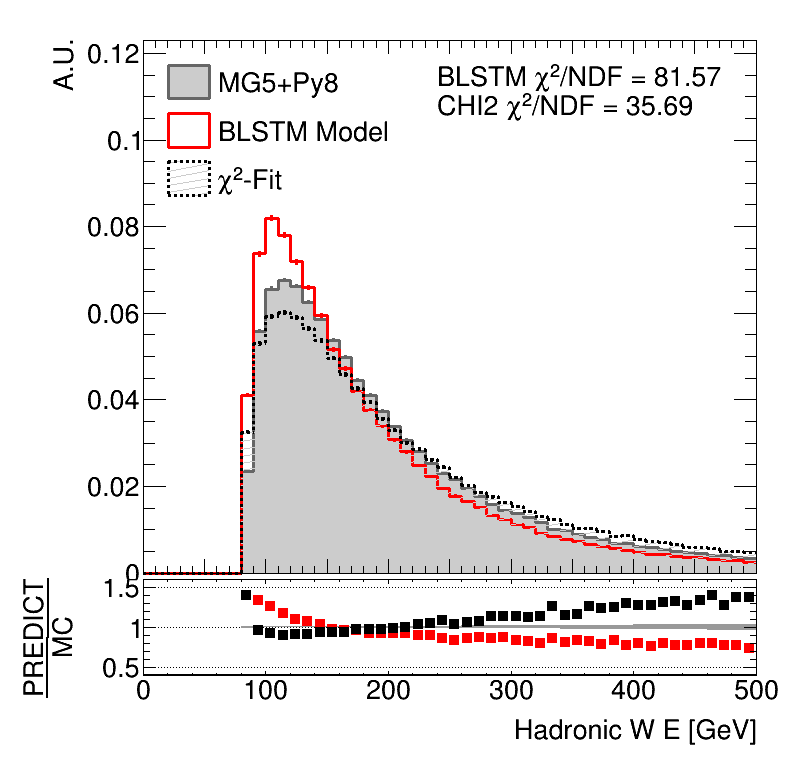}
\caption{\small{Reconstructed \Wboson-boson observables for the hadronic top quark. The gray filled area represents the  prediction obtained using the \MadGraph{}+\Pythia{}~Monte Carlo event generator. The black dashed line is obtained from the permutation of jets which minimizes the $\chi^2$. The red solid line is the output of the BLSTM.}}
\label{fig:results:W_had}
\end{figure}

\begin{figure}[htbp]
\centering
\includegraphics[scale=0.25]{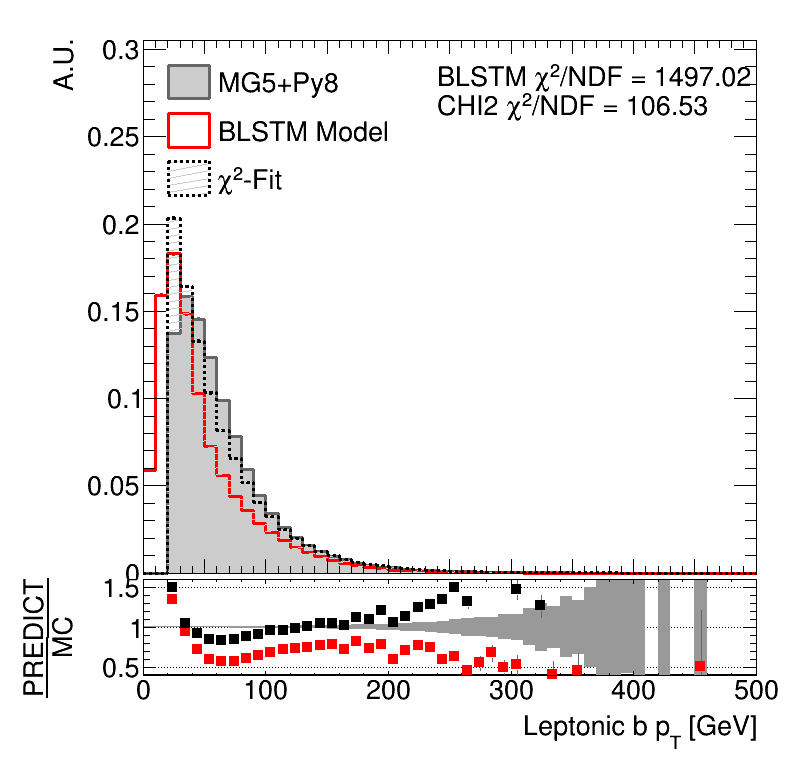}
\includegraphics[scale=0.25]{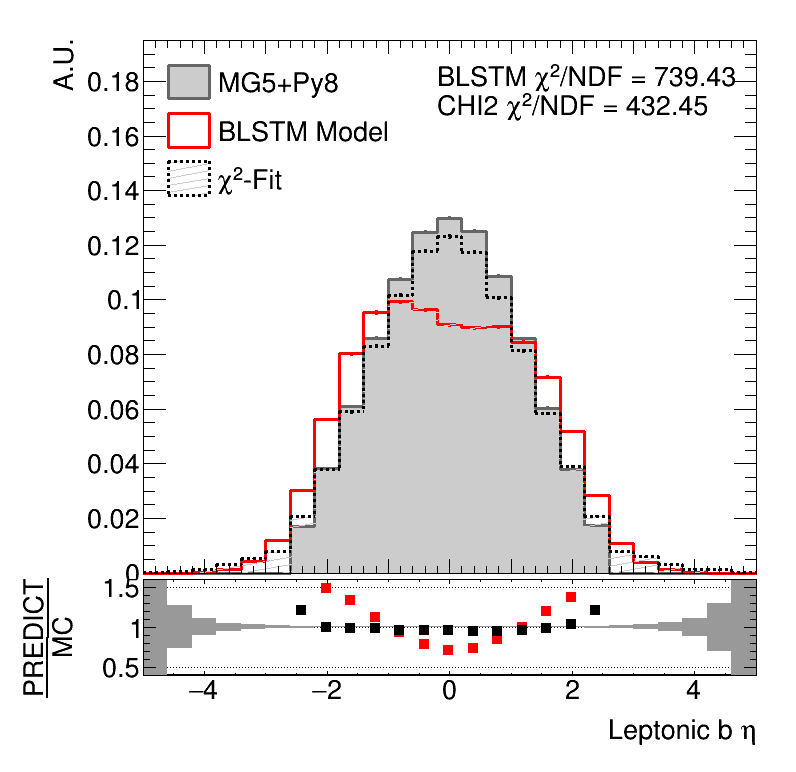}
\includegraphics[scale=0.25]{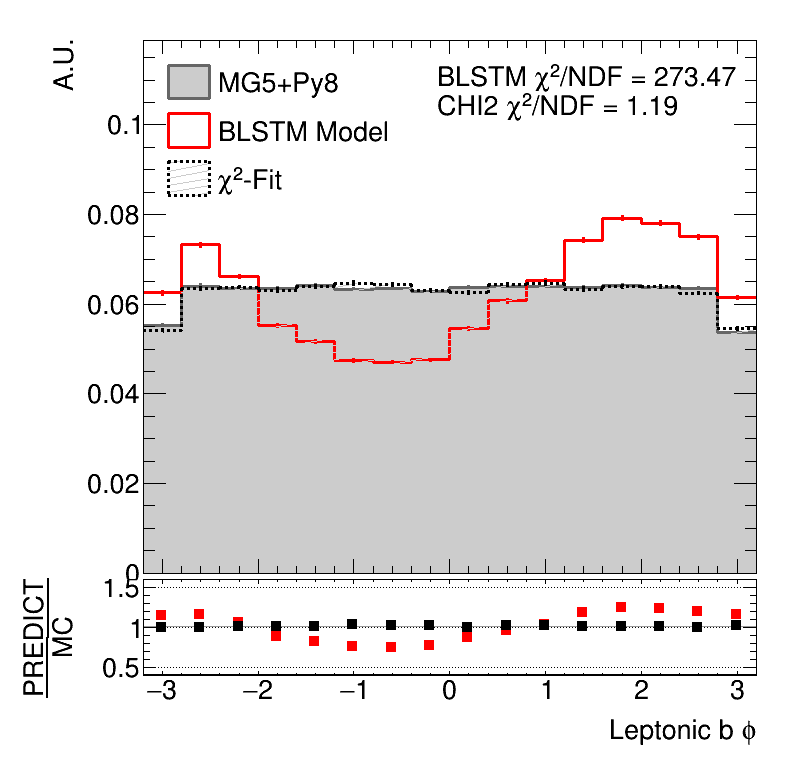}
\includegraphics[scale=0.25]{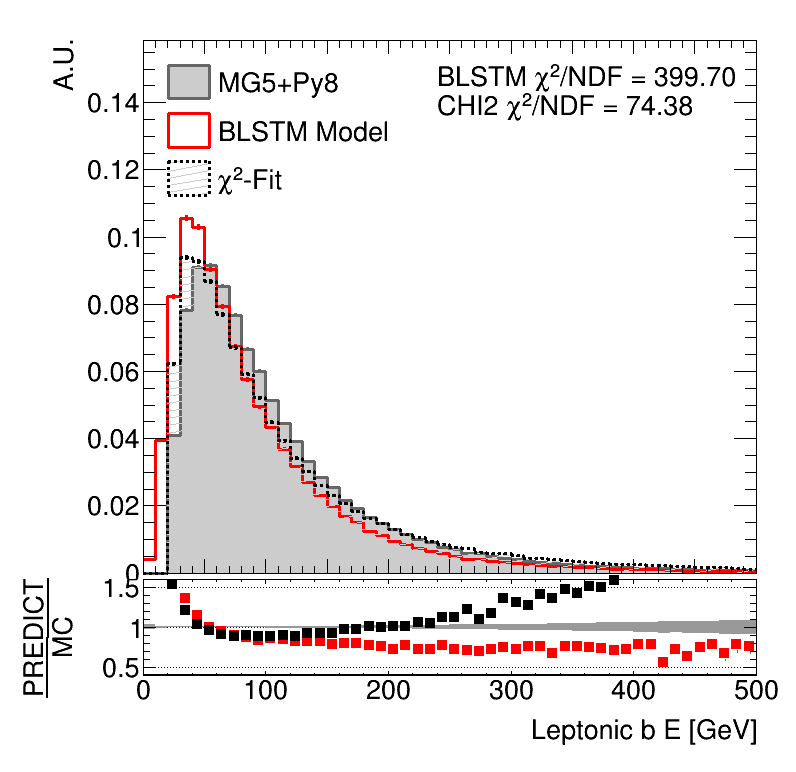}
\caption{\small{Reconstructed $b$-quark observables for the semileptonic top-quark. The gray filled area represents the  prediction obtained using the \MadGraph{}+\Pythia{}~Monte Carlo event generator. The black dashed line is obtained from the permutation of jets which minimizes the $\chi^2$. The red solid line is the output of the BLSTM.}}
\label{fig:results:b_lep}
\end{figure}

\begin{figure}[htbp]
\centering
\includegraphics[scale=0.25]{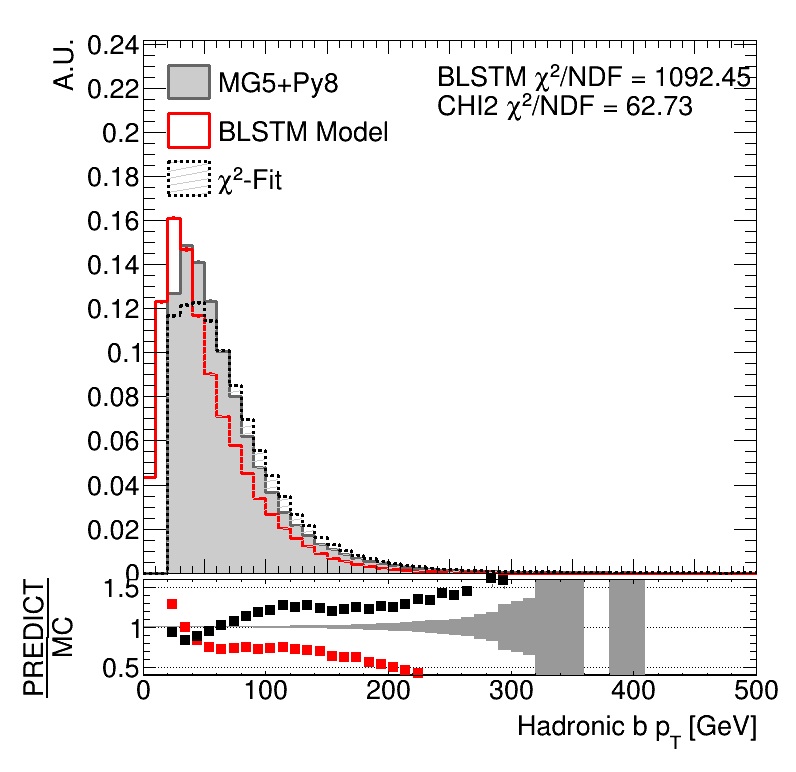}
\includegraphics[scale=0.25]{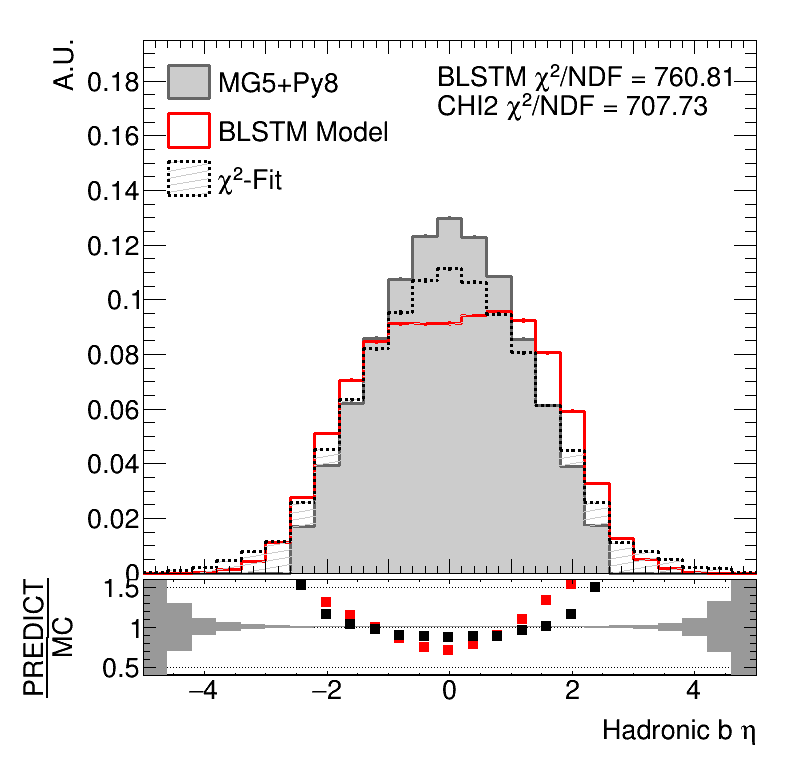}
\includegraphics[scale=0.25]{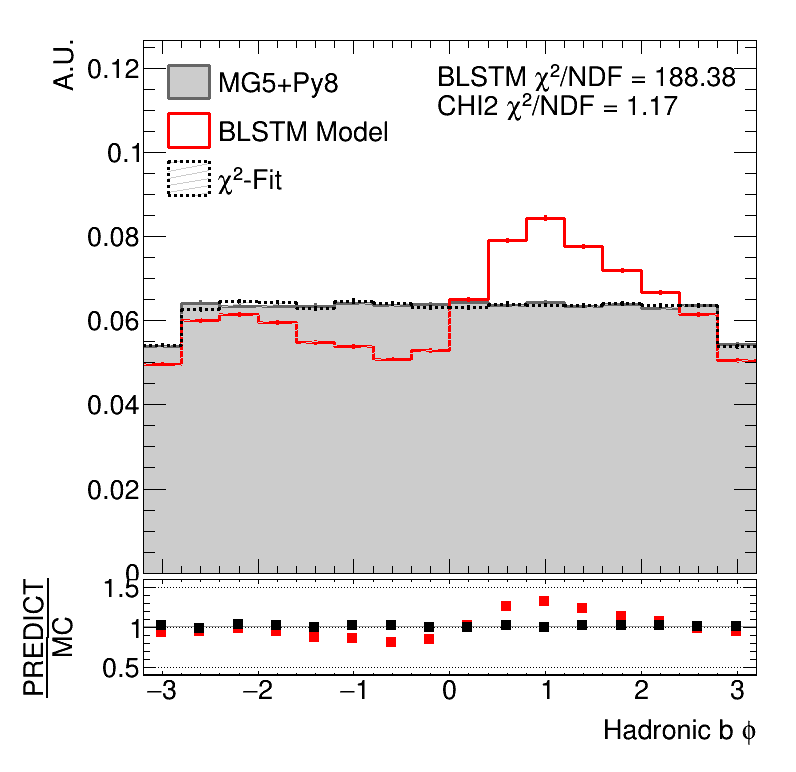}
\includegraphics[scale=0.25]{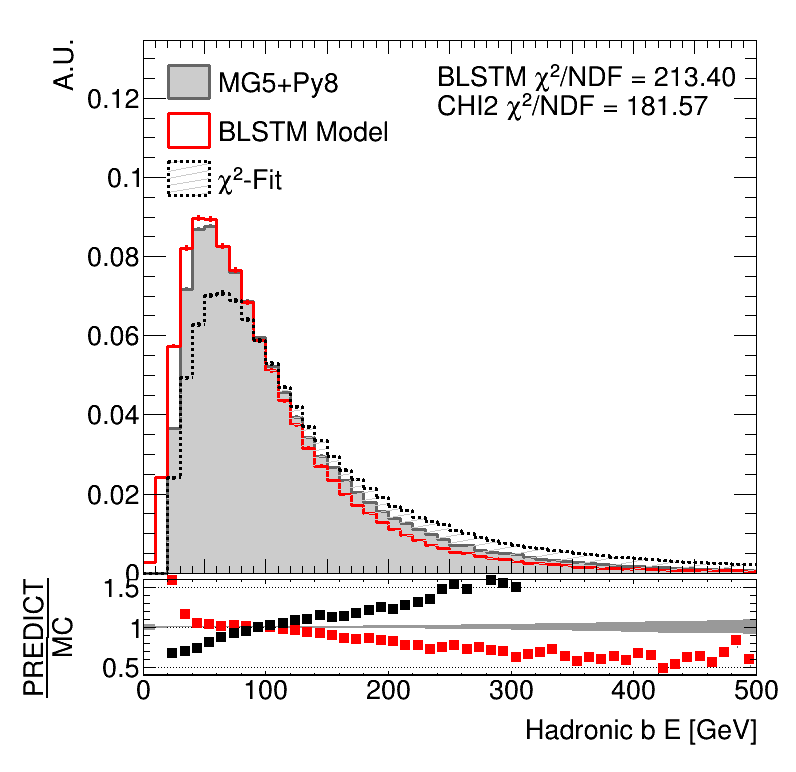}
\caption{\small{Reconstructed $b$-quark observables for the hadronic top quark. The gray filled area represents the  prediction obtained using the \MadGraph{}+\Pythia{}~Monte Carlo event generator. The black dashed line is obtained from the permutation of jets which minimizes the $\chi^2$. The red solid line is the output of the BLSTM.}}
\label{fig:results:b_had}
\end{figure}

Figures \ref{fig:results:t_lep} -- \ref{fig:results:b_had} show the reconstructed four-momenta of each of the six particles in the decay chain. We compare the normalized distributions predicted by \AngryTops{} and \chisqfit{} to those obtained using the MC event generator by the means of a $\chi^2$ metric defined as
\begin{eqnarray}
    \chi^2 / \textrm{NDF} &=& \frac{1}{n - 1} \sum_{i = 1}^{n} \frac{\left ( y_i^{\rm MC} - y_i^{\rm{predicted}} \right )^2}{\sigma_i^2} \\
{\textrm{where}}\quad    \sigma_i &=& \sqrt{\left (\sigma^{\rm MC}_i \right )^2 + \left (\sigma^{\rm predicted}_i \right )^2},
\end{eqnarray}
and where $y_i$ and $\sigma_i$ correspond to the value and uncertainty in the $i$-th histogram bin. The results of these comparisons are presented in Table~\ref{tab:agreement}.

\begin{table}[ht]
\begin{center}
 \begin{tabular}{c | c | c} 
 \hline
 \bf{Observable} & \multicolumn{2}{c}{\bf{$\chi^2$ / DOF}} \\
 \hline
   & \bf{BLSTM} & \bf{\chisqfit} \\ 
 \hline\hline
 $\pt^{t,lep}$ & 5.81  & 67.08 \\ 
 \hline
 $\eta^{t,lep}$ & 115.20 & 602.22 \\ 
 \hline
 $\phi^{t,lep}$ & 93.47  & 0.56  \\ 
 \hline\hline
 $\pt^{t,had}$ & 110.10 & 144.02 \\ 
 \hline
 $\eta^{t,had}$ & 16.28 & 471.56 \\ 
 \hline
 $\phi^{t,had}$ & 97.43 & 0.55 \\ 
 \hline\hline
  $\pt^{W,lep}$ & 156.94 & 142.99 \\ 
 \hline
 $\eta^{W,lep}$ & 162.24 & 729.31 \\ 
 \hline
 $\phi^{W,lep}$ & 35.30 & 0.66 \\ 
 \hline\hline
 $\pt^{W,had}$ & 498.41 & 270.30 \\ 
 \hline
 $\eta^{W,had}$ & 84.85 & 390.09 \\ 
 \hline
 $\phi^{W,had}$ & 174.08 & 1.16 \\ 
 \hline\hline
  $\pt^{b,lep}$ & 1497.02 & 106.53 \\ 
 \hline
 $\eta^{b,lep}$ & 739.43 & 432.45 \\ 
 \hline
 $\phi^{b,lep}$ & 273.47 & 1.19 \\ 
 \hline\hline
 $\pt^{b,had}$ & 1092.45  & 62.73 \\ 
 \hline
 $\eta^{b,had}$ & 760.81 & 707.73 \\ 
 \hline
 $\phi^{b,had}$ & 188.38 & 1.17 \\ 
 \hline\hline
\end{tabular}
 \caption{\small{Comparison of the difference in the distributions of the kinematic variables, as measured by $\chi^2$ / DOF between the results and the MC prediction, resulting from the BLSTM and \chisqfit{}\ reconstruction methods.}}
 \label{tab:agreement}
\end{center}
\end{table}

\newpage
\section{Observations} \label{Observations}
We first note that the $\chi^2$\ comparisons are not particularly
insightful, though they show some obvious differences in algorithm
performance.  In particular, the \chisqfit{}\ algorithm reproduces
the $\phi$~distributions well (which are expected to be featureless) while the \AngryTops{} algorithm typically creates some structure in
these distributions at the level of 10-20\%.  

The \AngryTops~algorithm shows a somewhat better performance on the semileptonic \Wboson-boson and the top quark kinematic variables as compared to the \chisqfit~matching fitter, as shown in Figures~\ref{fig:results:t_lep}\ and \ref{fig:results:W_lep}, but does not improve on the kinematics of $b$ quarks. 
We observe that the two models are generally closest in performance on the top quark kinematics, with the distributions of \chisqfit\ and \AngryTops{} matching closely with the MC distributions.
We note that the angular distributions are not particularly well reproduced by \AngryTops, a feature that appears to arise from either
incomplete training of the network, or a fundamental instability in how
these variables are reproduced by the BLSTM.  

Interestingly, both the neural network and the \chisqfit{} algorithms perform in a similar
manner with the hadronic top-quark kinematics.
In particular, both tend to under-estimate the top-quark \pt\ in the same manner.  \AngryTops{} predictions for the $\eta$~distribution are more accurate, though we see a remaining asymmetry in the $\phi$~distribution.

We observe that the kinematics of the \Wboson-boson and $b$-quark are 
reconstructed relatively poorly by both algorithms, a feature that 
is well-known
for the \chisqfit{}\ algorithm and is not improved by the BLSTM approach.
The $b$-quark kinematics are perhaps the most poorly reconstructed observables by \AngryTops{}, with a consistent under-estimation of the $b$-quark \pt~distribution.  
Although all $b$-quark jet candidates used in the training are required to have $\pt > 20$~\GeV, interestingly, \AngryTops{} predicts results that are below that threshold.

The choice of kinematic variables to represent the data also has a significant impact on the performance of the models. 
We find in general our models struggle most with learning the transverse momenta distributions. 
In all the particles besides the \Wboson-boson from the semileptonic top quark and the semileptonic top, \AngryTops~consistently underestimates the transverse momentum and even fails to learn the \pt~cutoff at 20~\GeV. This error arises from the under-estimate made on \px~and \py~by \AngryTops. 
The \chisqfit~on the other hand tends to slightly overestimate the transverse momentum, and while it fails to determine the cut-off in the $W$ boson transverse momenta, it is able to account for the cut-off in the $b$ quark transverse momenta.

There are also slight differences in the shape of the distributions between our models and the MC histograms. 
While \AngryTops~mostly learns the distributions, there are some asymmetries present that occasionally occur in the $\phi$ and rapidity distributions. These asymmetries are not a consistent phenomenon and differ between different training sessions. 
Due to the inherent complexity of \AngryTops, further studies of our  machine learning approach are necessary to better understand this behaviour. The \chisqfit~on the other hand does not present any asymmetries and significantly outperforms \AngryTops~in the $\phi$ variable. 

\section{Conclusions} \label{Conclusions}
In this study, we have analyzed the capability of using neural networks to perform kinematic reconstructions of particles in the semi-leptonic \ttbar~decay. 

While we do not claim to have the best network architecture for this problem, our work demonstrates the potential avenue for machine learning in this line of research. In Sections \ref{Results} and \ref{Observations}, we show the capability of machine-learning based approaches to be competitive with standard reconstruction algorithms such as \chisqfit.

The nature of a machine-learning based approach to kinematic reconstruction also presents additional advantages in improved flexibility. 
Algorithms such as \chisqfit, \KLFitter{} and \PseudoTop{} require fixed number of jets and inputs, while with \AngryTops{} one is easily able to update the input information of a network by adding/modifying network layers. 
An extension to the boosted regime, where quarks are collimated and appear as a single jet, seems straightforward with the current implementation based on recurrent neural networks. 
The major drawback of \AngryTops{} however is that one has little understanding of the intermediate steps performed by the network.

There are many ways to go beyond the \AngryTops{} project. 
Of course, the search for the ``best'' neural network architecture is an ongoing problem that can only be solved with further time and developments in machine learning techniques. 
Additionally, a comparison between \AngryTops{} and a more sophisticated kinematic reconstruction algorithm such as \KLFitter{} is necessary. 
We reserve this step of the analysis for a latter study, as differences in detector level simulations and input information make a direct comparison difficult given the complexity of these more sophisticated algorithms. 
It is possible that a combination of pre-existing reconstruction algorithms aided by machine-learning based approaches may also lead to significant advancements in the kinematic reconstruction of \ttbar{} final states.

\section*{Acknowledgments}
We acknowledge the support of the Natural Sciences and Engineering Research Council of Canada (NSERC).

\FloatBarrier

\bibliographystyle{IEEEtran}
\bibliography{main.bib}

\end{document}